# Fiber-distributed Signal Processing: Where the Space Dimension Comes into Play

Ivana Gasulla, Sergi García, David Barrera, Javier Hervás and Salvador Sales
*ITEAM Research Institute, Universitat Politècnica de València, 46022 Valencia, Spain*
*Author e-mail address: ivgames@iteam.upv.es*

**Abstract:** We present how to implement "fiber-distributed signal processing" in the context of fiber-wireless communications by using different multicore fibers, providing both radio access distribution and microwave photonics signal processing in the same fiber medium.
**OCIS codes:** (060.2330) Fiber optics communications; (060.5625) Radio frequency photonics; (060.3735) Fiber Bragg gratings

## 1. Introduction

Space-division multiplexing (SDM) fibers were originally envisioned for high-capacity digital communications in the context of core and metro networks, [1]. Nevertheless, evolving fiber-wireless access networks could greatly benefit form SDM technologies as well since emerging communications scenarios, as 5G and Internet of Things, demand on higher degrees of capacity, connectivity, flexibility and compactness, [2]. In this context, we have proposed the use of multicore fibers (MCFs) as a compact medium to implement "fiber-distributed signal processing", i.e., a medium that provides -simultaneously- both radio access signal distribution (including MIMO antenna connectivity) and radiofrequency (RF) signal processing [3,4]. Many of the required signal processing functionalities, such as microwave filtering, radio beam-steering and signal generation, rely on a key building block: the true time delay line (TTDL), [5]. Figure 1 shows a representative fiber-wireless communications scenario where the TTDL is required as both a fiber-distributed processing link and a compact processing device. The inset of Fig. 1 shows the general idea underlying the MCF-based TTDL, where we obtain a set of different time-delayed samples of the RF signal characterized by a constant basic differential delay $\Delta\tau$ between adjacent samples.

We show in this paper how the introduction of the space dimension brings the implementation of fiber-distributed signal processing following two different approaches: compact multicavity processing elements using commercial homogeneous MCFs and fiber processing links using custom designed heterogeneous MCFs.

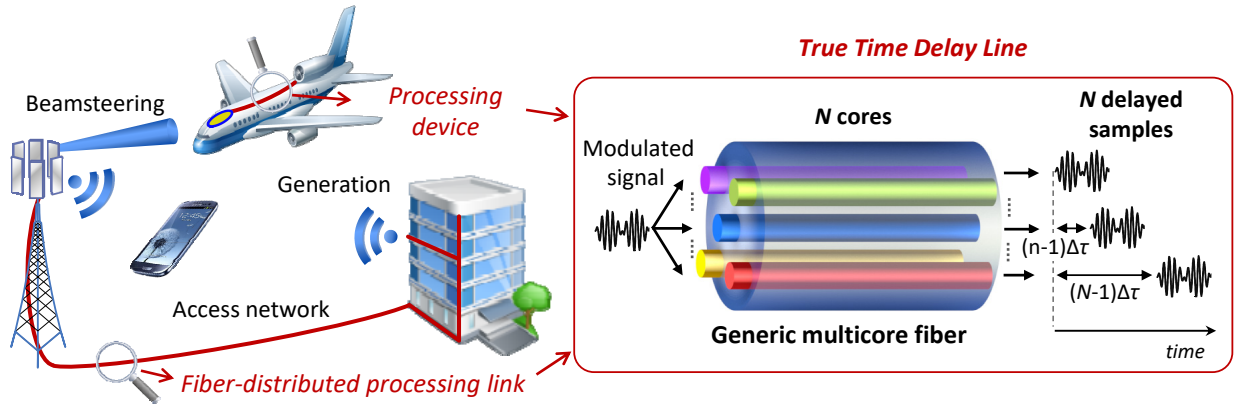

Fig. 1. Application fiber-wireless communications scenario and sampled true time delay line built upon a generic multicore fiber

## 2. How to implement true time delay lines on different multicore fiber technologies

### 2.1. Compact multicavity devices built upon homogeneous multicore fibers

Homogeneous MCFs comprise a set of cores that are -in principle- identical and share the same group delay characteristics. Hence, a novel approach for designing true time delay lines based on commercial homogeneous MCFs requires the flexible in-line incorporation of additional dispersive elements, such as fiber Bragg gratings (FBGs) in each core. This involves the design of a multicavity device where the diversity in the propagation characteristics comes from the accurate inscription of FBGs at selective positions along the cores. Previous work on FBG inscription in MCFs has focused mainly on the simultaneous writing of the same grating (i.e., placed at the same position) in all the cores, mainly for sensing and astrophotonics. In order to inscribe gratings selectively, we have developed the first fabrication method to inscribe high-quality gratings characterized by arbitrary frequency spectra and located in arbitrary longitudinal positions along the individual cores, [3]. The inscription procedure is based on the moving phase mask technique. The beam is conditioned by a mirror mounted in a piezo-electric transducer that can induce a beam vertical deflection and by two cylindrical lenses that control the beam height and width. Actually, the dimension and cross-sectional arrangement of the cores determine the height, which was confined to a range from 30 to 50 µm. We adjust the beam width and the distance of the MCF to the phase mask to control the interference pattern created after the phase mask. For a phase mask with a period of 1070 nm, the maximum width that allows single core inscription in one outer core is approximately 23 µm.

Our group has fabricated different multicavity TTDL devices built upon the inscription of FBGs in homogenous MCFs, [3]. Fig. 2(a) shows the scheme of a multicavity device that is based on the selective inscription of FBGs in 3 outer cores (labeled as #4, #5 and #6). In each of these cores, we inscribed an array of three equally-spaced uniform gratings, being each one located in a different longitudinal position and centered at a different Bragg wavelength. The MCF is a commercial 7-core fiber with a cladding diameter of 125 µm and a core pitch of 35 µm, (from Fibercore). To assure a proper wavelength diversity operation, the 3 arrays of FBGs have been written to feature incremental distances between FBGs from core to core, i.e., 20-mm, 21-mm and 22-mm distances, respectively, within cores 6, 5 and 4. On the other hand, the spatial diversity arises from the incremental displacement applied between FBGs centered at the same wavelength but inscribed in adjacent cores, i.e., 6-mm, 7-mm and 8-mm displacements, respectively, for the wavelengths $\lambda_1$ = 1537.07 nm, $\lambda_2$ = 1541.51 nm and $\lambda_3$ = 1546.26 nm.

We experimentally evaluated the performance of the fabricated multicavity TTDL device in the context of reconfigurable microwave signal filtering, [5]. The filtering effect is achieved by combining and photodetecting all the delayed RF signal samples coming from the MCF output. The Free Spectral Range (FSR) of the filter is then given by the inverse of the basic differential delay $\Delta\tau$ obtained between adjacent samples. Figure 2(b) shows the measured electrical frequency responses for the 3-tap filters obtained when we operate in wavelength diversity, i.e. gathering the delayed samples coming from a given single core (cores 4, 5 or 6). We see how we can reconfigure the filter, varying the FSR from 4.45 up to 4.97 GHz. The same device can operate using the spatial diversity provided by the cores, i.e. gathering the delayed samples coming from different cores that correspond to the same single optical wavelength ($\lambda_1$, $\lambda_2$ or $\lambda_3$). Figure 2(c) shows the 3-tap filter responses given by the three input optical wavelengths, where we can see how the FSR varies from 12.50 up to 17.76 GHz. These experimental results are in good agreement with the theoretical ones, with small discrepancies in the basic differential delays that may be caused by discrepancies between the theoretical and actual values of the core refractive indices and small imbalances in the reflectivity strength of the FBGs.

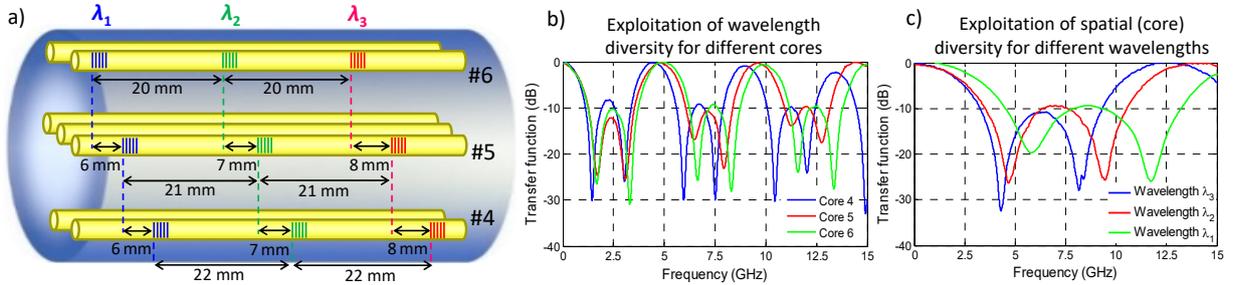

Fig. 2. (a) Schematic of the arrays of FBGs selectively inscribed in cores 4, 5 and 6; (b) Measured 3-tap filter transfer functions using wavelength diversity (for each one of the target cores) and (c) Measured 3-tap filter transfer functions using spatial diversity (for each one of the wavelengths).

## 2.2. Fiber-distributed processing links built upon heterogeneous multicore fibers

If we consider that the refractive index profile of each core can be designed independently, we can implement MCFs that operate as group-index-variable delay lines. This gives us the possibility of implementing fiber-distributed signal processing links with a length up to a few kilometers. A sampled TTDL built upon a heterogeneous MCF implies that each core $n$ features an independent group delay with a linear dependence on the optical wavelength. The spectral group delay $\tau_{g,n}(\lambda)$ of a core $n$ can be expanded as a 3$^{rd}$-order Taylor series around an anchor wavelength $\lambda_0$ as

$$\tau_{g,n}(\lambda) = \tau_g(\lambda_0) + D_n \cdot (\lambda - \lambda_0) + \frac{S_n}{2} \cdot (\lambda - \lambda_0)^2, \tag{1}$$

where $D_n$ is the chromatic dispersion and $S_n$ the dispersion slope of core $n$. For proper operability [4], we need that: $\tau_{g,n}(\lambda_0)$ is common to all cores, $D_n$ increases linearly with $n$ and a linear behavior of the $\tau_{g,n}$ within the optical operation range. In addition, we must assure lower levels of intercore crosstalk and robustness against fiber bends. If we focus, for instance, on the spatial diversity regime, the basic differential delay between samples is given by the propagation difference created between two adjacent cores for a particular wavelength $\lambda_m$, that is $\Delta\tau = \tau_{g,n+1} - \tau_{g,n} = \Delta D \cdot (\lambda_m - \lambda_0) + (S_{n+1} - S_n)/2 \cdot (\lambda_m - \lambda_0)^2$. Here, $\Delta D = D_{n+1} - D_n$ is the incremental dispersion between adjacent cores that must be the same for each pair of adjacent cores. The design of the TTDL must as well minimize the dispersion slope variation $S_{n+1} - S_n$ to reduce the group delay quadratic wavelength dependence.

In previous work, e.g. [4], we have designed and evaluated different heterogeneous MCF links attending to different delay line and transmission performance requirements. These fibers are based on 7-core structures with different trench-assisted refractive index profiles that share a 35-μm core pitch and a 125-μm cladding diameter. Figure 3(a) shows the cross section of one of the MCFs designed where each core is characterized by different parameters that vary within the following ranges: core radius $a_1$ from 3.42 up to 4.98 μm, core-to-cladding relative index difference $\Delta_1$ from 0.3333% up to 0.3864%, core-to-trench separation $a_2$ from 2.42 up to 5.48 μm and trench width $w$ from 2.61 up to 5.41 um. The cladding-to-trench relative index is $\Delta_2 \approx 1\%$ in all cores. The fiber satisfies the common group index condition at $\lambda_0 = 1550$ nm as well as the common incremental dispersion, $\Delta D = 1$ ps/(km·nm), with a set of dispersion parameters $D_n$ ranging from 14.75 up to 20.75 ps/(km·nm). Figure 3(b) shows the computed group delay for each core as a function of the optical wavelength, where we observe linearly incremental group delay slopes for each core.

As a proof of concept, we evaluated as well the performance of the designed TTDL when it is applied to reconfigurable microwave signal filtering. Figure 3(c) shows the computed transfer function of the microwave filter as a function of the RF frequency for a 5-km MCF link and two different optical wavelengths. We can see how the increase of the optical wavelength from 1560 to 1570 nm reduces the FSR from 20 to 10 GHz without any filter degradation due to higher-order dispersion effects. Other microwave photonics functionalities, as optical beamforming for phased array antennas, were also evaluated in [4].

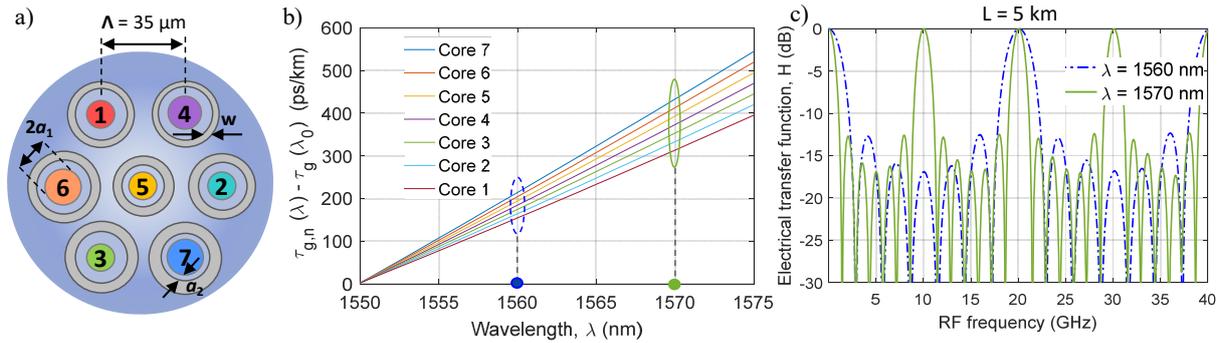

Fig. 3. (a) Cross section of a heterogeneous MCF designed with different refractive index profiles; (b) Computed group delay versus optical wavelength for all the cores and (c) Computed transfer function of the microwave signal filter for a 5-km link and wavelengths 1560 and 1570 nm.

## 4. Conclusions

We have shown how the space dimension brings the implementation of distributed RF signal processing in a single fiber following two different approaches: compact multicavity elements using commercial homogeneous MCFs and fiber links using custom designed heterogeneous MCFs. These solutions cover a variety of application scenarios that will be specially demanded in 5G communications providing, among others, arbitrary waveform generation, microwave signal filtering, optical beamforming networks for phased array antennas and MIMO antenna connectivity.

**Acknowledgements**

This research was supported by the ERC Consolidator Grant 724663, the Spanish Projects TEC2014-60378-C2-1-R, TEC2015-62520-ERC and TEC2016-80150-R, the Spanish scholarships MECD FPU13/04675 for J. Hervás and MINECO BES-2015-073359 for S. García, and Spanish MINECO Ramón y Cajal RYC-2014-16247 for I. Gasulla.